\documentclass[%
 reprint,
superscriptaddress,
bibnotes,
 amsmath,amssymb,
 aps,prl,
]{revtex4-1}
\usepackage{graphicx}
\usepackage{dcolumn}
\usepackage{bm}

\begin{document}

\preprint{}

\title{Fickian Yet non-Gaussian Diffusion in Complex Molecular Fluids via a non-local diffusion framework}

\author{H. Srinivasan}
\email{harish.snvsn@gmail.com}
\affiliation{Solid State Physics Division, Bhabha Atomic Research Centre, Mumbai 400085, India}
\affiliation{Homi Bhabha National Institute, Anushaktinagar, Mumbai 400094, India}
 
\author{V. K. Sharma}
\affiliation{Solid State Physics Division, Bhabha Atomic Research Centre, Mumbai 400085, India}
\affiliation{Homi Bhabha National Institute, Anushaktinagar, Mumbai 400094, India}

\author{S. Mitra}
\email{smitra@barc.gov.in}
\affiliation{Solid State Physics Division, Bhabha Atomic Research Centre, Mumbai 400085, India}
\affiliation{Homi Bhabha National Institute, Anushaktinagar, Mumbai 400094, India}

\begin{abstract}
Fickian yet non-Gaussian diffusion (FnGD) has gained popularity in the recent times owing to it’s ubiquity in a variety of complex fluids. However, whether FnGD can be observed experimentally in molecular fluids is still obscure with very little study in real systems. In this letter, we show existence of FnGD in molecular fluids based on compelling evidence from incoherent quasielastic neutron scattering (IQENS). Using a cage-jump diffusion model, we show that while the approach to Fickianity is exponentially fast, the Gaussianity is restored at a much slower algebraic rate. We propose a non-local diffusion (NLD) model to describe a $d$-dimensional jump-diffusion in FnGD regime and show their universal applicability in such systems. This study establishes that cage-jump diffusion process inevitably lead to FnGD and provides the framework of NLD models to explore such diffusion phenomena in any arbitrary dimensions.

\end{abstract}

\maketitle


The distinguishing characteristic of Brownian motion is the simultaneous presence of a Gaussian distribution of particle displacements and a mean-squared displacement (MSD) that varies linearly with time \cite{Einstein_1905}. However, anomalous diffusion phenomena challenge this behavior, exhibiting non-linear variations in MSD accompanied by non-Gaussian/Gaussian displacement distributions \cite{Balakrishnan_1985, Metzler_2000, Meroz_2015, Wei_2022}. Interestingly, while non-linear MSD with a Gaussian distribution has been observed in various systems, the contrary behaviour of linear MSD with non-Gaussian distribution had remained elusive until the recent discovery by Granick's group \cite{Wang_2009, Wang_2012}. They observed  the intriguing occurrence of linear MSD despite non-Gaussian distribution of displacements in certain colloidal systems.  This discovery led to the development of new models, known as Fickian yet non-Gaussian diffusion (FnGD), which aimed to explain this phenomenon in terms of structural and dynamical heterogeneities \cite{Wang_2012,Chubynsky_2014,Metzler_2017,Chechkin_2017,Jain_2016,Kim_2013,Guan_2014, Matse_2017, Bialas_2020, Pastore_2021, Wang_2009, Hapca_2008, Jain_2017a}.

Drawing inspiration from superstatistics\cite{Wang_2009, Hapca_2008} and subordination\cite{Chechkin_2017, Jain_2016, Jain_2017}, these models captured the emergence of FnGD from the stochastic nature of the system's environment. Recent studies \cite{Pastore_2021,Rusciano_2022} demonstrated the existence of FnGD in 2D colloidal glass-formers and established a correlation between FnGD and the system's dynamical heterogeneity, indicating that FnGD was enhanced in the neighbourhood of glass-transition. At the core of such systems lies the mechanism of cage-jump diffusion processes \cite{Pastore_2014, Bier_2008, Vorselaars_2007, PicaCiamarra_2016, Candelier_2010}, prompting the question of whether FnGD can be attributed to this mechanism. Specifically, can molecular systems, which exhibit cage-jump diffusion, also perhaps display FnGD? Further, can the nature of cage-jump mechanism also be described as a source of environmental stochasticity in the diffusion?

In this letter, we address these questions through a two-fold approach. First, we provide compelling evidence of FnGD in a range of molecular liquids based on incoherent quasielastic neutron scattering (IQENS) experiments. Second, we develop a physically intuitive model for non-local diffusion (NLD) that effectively captures FnGD regime. The subordination technique is used to solve the observed jump-diffusion model in FnGD regime for any $d$-dimensional system. These findings expand the scope of rapidly evolving FnGD models to molecular liquids and offer a physical foundation for understanding this behavior in systems characterized by cage-jump diffusion through NLD models.

\textbf{\textit{Cage-jump molecular diffusion}}

Molecular self-diffusion in various complex fluids is characterized through transient caging followed by jump diffusion, which arises from a dynamic equilibrium of transient structures formed by intermolecular hydrogen bonding or ionic complexation. This behaviour is observed in a range of complex liquids including supercooled water \cite{Qvist_2011}, ionic liquids \cite{Embs_2012, Aoun_2013, Burankova_2014}, deep eutectic solvents \cite{Wagle_2015,Srinivasan_2020}. IQENS is a suitable technique to probe the diffusion landscape at molecular length and time scales, providing comprehensive insights into the nature of the diffusion process and a direct link with van-Hove self-correlation functions, owing to its spatiotemporal sensitivity \cite{Bee_1988}. In this study, we collate IQENS findings from multiple sources \cite{Qvist_2011, Embs_2013, Aoun_2013, Burankova_2014, Wagle_2015, Srinivasan_2020} on different complex fluids, which have been successfully described using a two-component model based on localized caged diffusion and jump diffusion.

Typically, the IQENS spectra, $S(Q, \omega)$, is given as a function of momentum transfer, $Q$, and energy transfer, $\omega = E/\hbar$. In systems executing cage-jump diffusion, the $S(Q,\omega)$ is given as a convolution of IQENS spectra corresponding to caged and jump diffusion processes \cite{SM},
\begin{equation}
\label{qens model}
\resizebox{0.9\hsize}{!}{$S(Q,\omega) = L_j(\Gamma_j,\omega) \otimes \Big[A_0 \delta(\omega) + \left(1-A_0\right) L_{loc}(\Gamma_{loc}, \omega) \Big]$}
\end{equation}
where $L_j(\Gamma_j,\omega)$ and $L_{loc}(\Gamma_{loc},\omega)$ are the Lorentzians associated to jump and caged diffusion processes. Here, $\Gamma_j(Q)$ and $\Gamma_{loc}(Q)$ are the half-width at half maximum (HWHM) of the respective Lorentzians, and correspond to $Q$-dependent relaxation rates associated to the jump and localized caged diffusion processes, respectively. $A_0(Q)$ is the Fourier transform of steady state distribution of the caged-diffusion process (it is also referred to as the elastic incoherent structure factor (EISF)\cite{Bee_1988}). IQENS data of numerous complex fluids have been shown to follow eq. \eqref{qens model} \cite{Qvist_2011, Embs_2012, Aoun_2013, Burankova_2014, Wagle_2015, Srinivasan_2020, Srinivasan_2023}, which essentially corresponds to a sum of two Lorentzians. The details of the development of the model describing cage-jump diffusion has been expounded in the supplementary material (SM)\cite{SM}. In what follows, we discuss the behaviour of parameters in eq. \eqref{qens model} which characterize the nature of the diffusion process.

\begin{figure}
	\centering
	\includegraphics[width=0.8\linewidth]{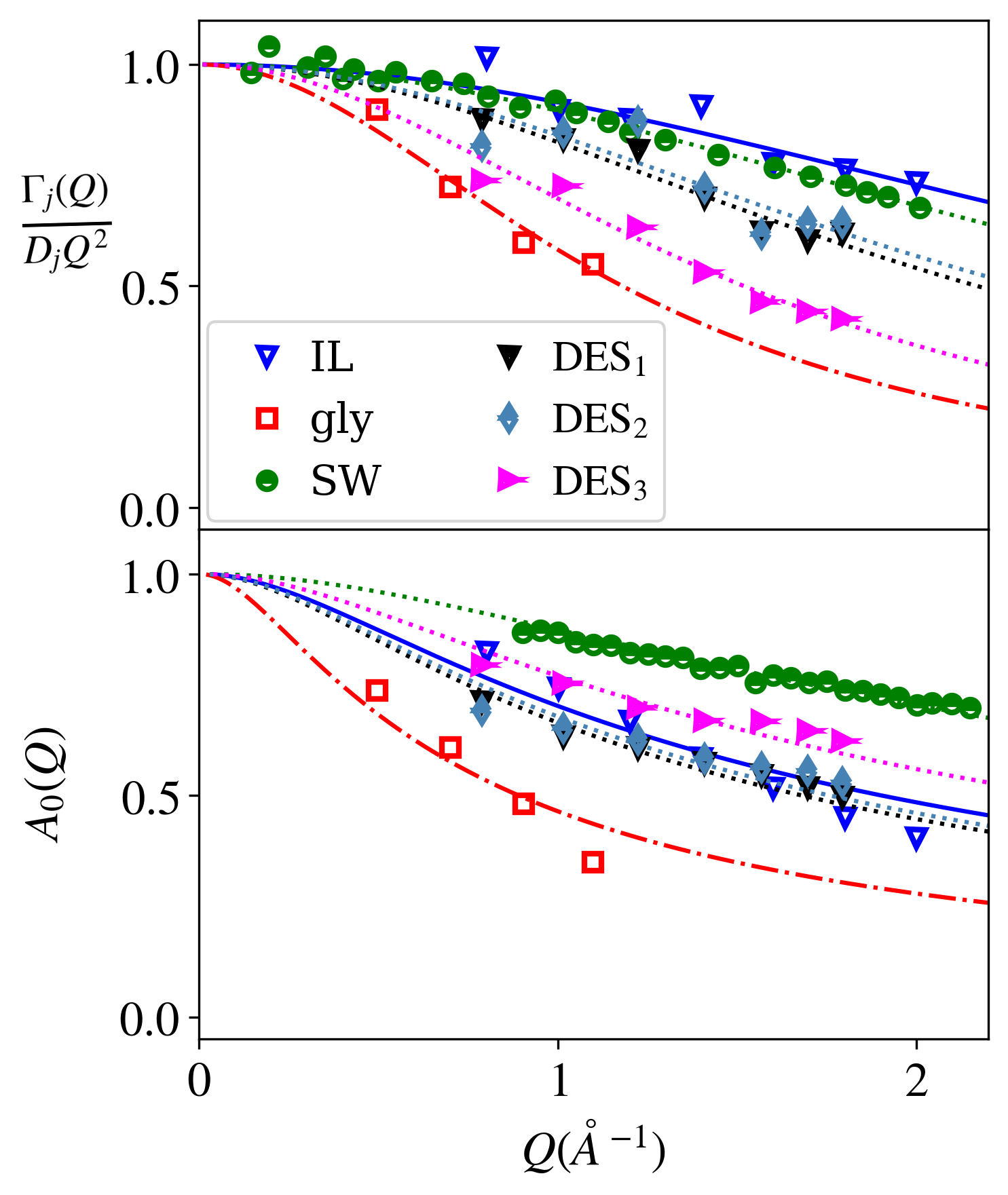}
	\caption{(top) The variation of normalized relaxation rate $\frac{\Gamma_j(Q)}{D_jQ^2}$ with respect to $Q$ for different systems surveyed in this study (IL: ionic liquid; gly: Glycerol; SW: Supercooled water; $\text{DES}_i$: Deep eutectic solvents. The fits based on jump diffusion model is also shown along side IQENS data points. (bottom) Elastic incoherent structure factor (EISF) of these systems obtained through IQENS experiments. The fits are based on soft-confinement with radii varying through an exponential distribution is also shown.}
\end{figure}

The jump diffusion process is generally described by an expression derived for a two-state diffusion model \cite{Singwi_1960}, $\Gamma_j (Q) = D_j Q^2 \Big[1+\tau_j D_j Q^2 \Big]^{-1}$, where $D_j$ is the jump diffusion constant and $\tau_j$ is the mean-waiting time between jumps. It is clear that at low-$Q$ limit, $\Gamma_j \sim D_j Q^2$, which corresponds to the Gaussian limit. Notably, the parameters $D_j$ and $\tau_j$ are also related to the characteristic mean-squared jump length given by, $l_0^2 = D_j \tau_j$.  Figure 1(a) presents the variation of $\Gamma_j/(D_jQ^2)$ with $Q$ for various systems. At low $Q$-values, the $\Gamma_j/(D_jQ^2)$ approaches 1, indicating that the diffusion behaviour approaches Gaussian limit at large distances. The deviation from 1, characterizes the strength of non-Gaussian behaviour in the system. Evidently, the jump diffusion of molecules exhibits strongly non-Gaussian behaviour at short-distance (high $Q$) and Gaussian behaviour at long-distances (low $Q$). The model's excellent fits with parameters $D_j$ and $\tau_j$ for various molecular fluids, including DESs studied in this work, indicate the universality of the underlying diffusion mechanism. It is notable that the model is robust and reliable despite the diverse chemical nature and complex structure of the fluids .

The $Q$-dependence of $A_0(Q)$ and $\Gamma_{loc}(Q)$ comprise information about the nature of the caged diffusion process. They are typically described using localized diffusion within an isotropic confinement \cite{Volino_1980,Volino_2006}. Here, we consider a soft-confinment in a spherical cage, whose radius is considered to be exponentially distributed. As shown in SM \cite{SM}, $A_0(Q)$ depends on the average radius $\sigma_0$, according to
\begin{equation}
A_0(Q) = \frac{\sqrt{\pi}}{2Q\sigma_0} e^{ \left[\frac{1}{4Q^2\sigma_0^2}\right]} \sqrt{\pi}\mathrm{Erfc}\left[\frac{1}{2Q\sigma_0}\right]
\end{equation}
This model shows excellent compatibility with the data for the wide range of systems studied, as demonstrated in Fig. 1 (b). The decay of $A_0$ to a lower value indicates a higher average caging radius $\sigma_0$. Combining these fits with  the description of $\Gamma_{loc}(Q)$ \cite{SM}, which provides information about typical timescale of diffusion within cages, $\tau_0$, it can be noted that the soft-confinement model serves as an excellent candidate for describing caged dynamics of molecules in these complex fluids. The complete description of IQENS using the cage-jump diffusion model is captured through these four important parameters $- \;\tau_j, l_0, \tau_0$ and $\sigma_0$, wherein the former two parameters describe the jump-diffusion process and the latter correspond to caged-diffusion.

In order to explore the emergence of FnGD in these systems, we would like to calculate two quantities\cite{SM}: non-Fickian parameter (NFP) and non-Gaussian parameter (NGP) which describe the deviation of the system from Fickian and Gaussian behaviour respectively. Both NFP \& NGP are essentially linked to the moments of the displacement, which can be directly calculated from the self-intermediate scattering function (SISF), $I(Q, t)$, through, $\left< \delta r^n(t) \right> =  (i \nabla_\mathbf{Q})^n I(Q,t)|_{Q=0}$. In systems which follow cage-jump diffusion mechanism, $I(Q,t)=e^{-\Gamma_j (Q)t} \left[A_0 (Q)+ \left(1-A_0 (Q)\right) e^{-\Gamma_{loc} (Q)t} \right]$ (Fourier transform of $S(Q,E)$ in eq. \eqref{qens model}). For the cage-jump diffusion model, it follows \cite{SM} that NFP, $\mu(t)$, and NGP, $\alpha_2(t)$, is,
\begin{equation}
\begin{gathered}
\label{nfp and ngp}
\mu(t) = \frac{2\sigma_0^2 }{l_0^2} \frac{\tau_j}{\tau_0} e^{-t/\tau_0} \\
\alpha_2(t) = 2\frac{\tau_j}{t} \left< \left[1 + \frac{\sigma^2}{l_0^2} \frac{\tau_j}{t}   \left(1-e^{-t/\tau_0}\right) \right]^{-2} \right>_\sigma
\end{gathered}
\end{equation}
Plugging in the values of $\sigma_0$, $l_0$, $\tau_j$ and $\tau_0$ are obtained from the model fits of IQENS spectra into eq. \eqref{nfp and ngp}, we plot the  NFP and NGP for various complex fluids in Fig. 2. Evidently, $\mu(t)$ shows a rapid-decay to zero, whereas $\alpha_2(t)$ persists over longer duration for all the systems, establishing the existence of a Fickian non-Gaussian diffusion (FnGD) regime. The crossover of $\alpha_2(t)$ and $\mu(t)$ is notably observed in the window of $2 - 5\; \tau_j$, suggesting that in this time-window the system has achieved the Fickian regime, while persists to exhibit non-Gaussian behaviour.

The FnGD regime can be more lucidly shown by coarsening out the caged-diffusion component. As $\tau_0$ is the relaxation timescale of caged motion, it is clear that always $\tau_j > \tau_0$ (as observed for all the complex fluids \cite{SM}). Therefore, in the limit $t \gg \tau_0$, we note that $\mu(t) \rightarrow 0$ indicating a Fickian behaviour, whereas $\alpha_2(t) \rightarrow \alpha_2^{FnG}(t)$
\begin{equation}
\alpha_2^{FnG}(t) = 2 \frac{\tau_j}{t}\left< \left[1+ \frac{\sigma^2}{l_0^2} \frac{\tau_j}{t}\right]^{-2} \right>_\sigma
\end{equation} 
Asymptotically, the leading order of decay is governed by $\alpha_2^{FnG}(t) \sim (t/\tau_j)^{-1}$. This long-time universal behaviour is also evident from Fig. 2, as $\alpha_2(t)$ for all the system falls on to a single master-curve. These analyses clearly establish that the existence of an FnGD regime which is marked by the limit $t \gg \tau_0$. Further, it can be noted that the SISF in the FnGD limit follows \cite{SM},
\begin{equation}
\label{iqt fngd}
I_{FnG}(Q,t)= A_0(Q) \exp\left[ \frac{-(Q l_0)^2}{1 + (Q l_0)^2} \frac{t}{\tau_j} \right]
\end{equation}
which clearly highlights that while the relaxation function decays exponentially, the spatial dependence is strongly non-Gaussian. In the next section, a non-local diffusion (NLD) model is proposed that captures the behaviour of $I_{FnG}(Q,t)$.
\begin{figure}
\centering
    \includegraphics[width=0.8\linewidth]{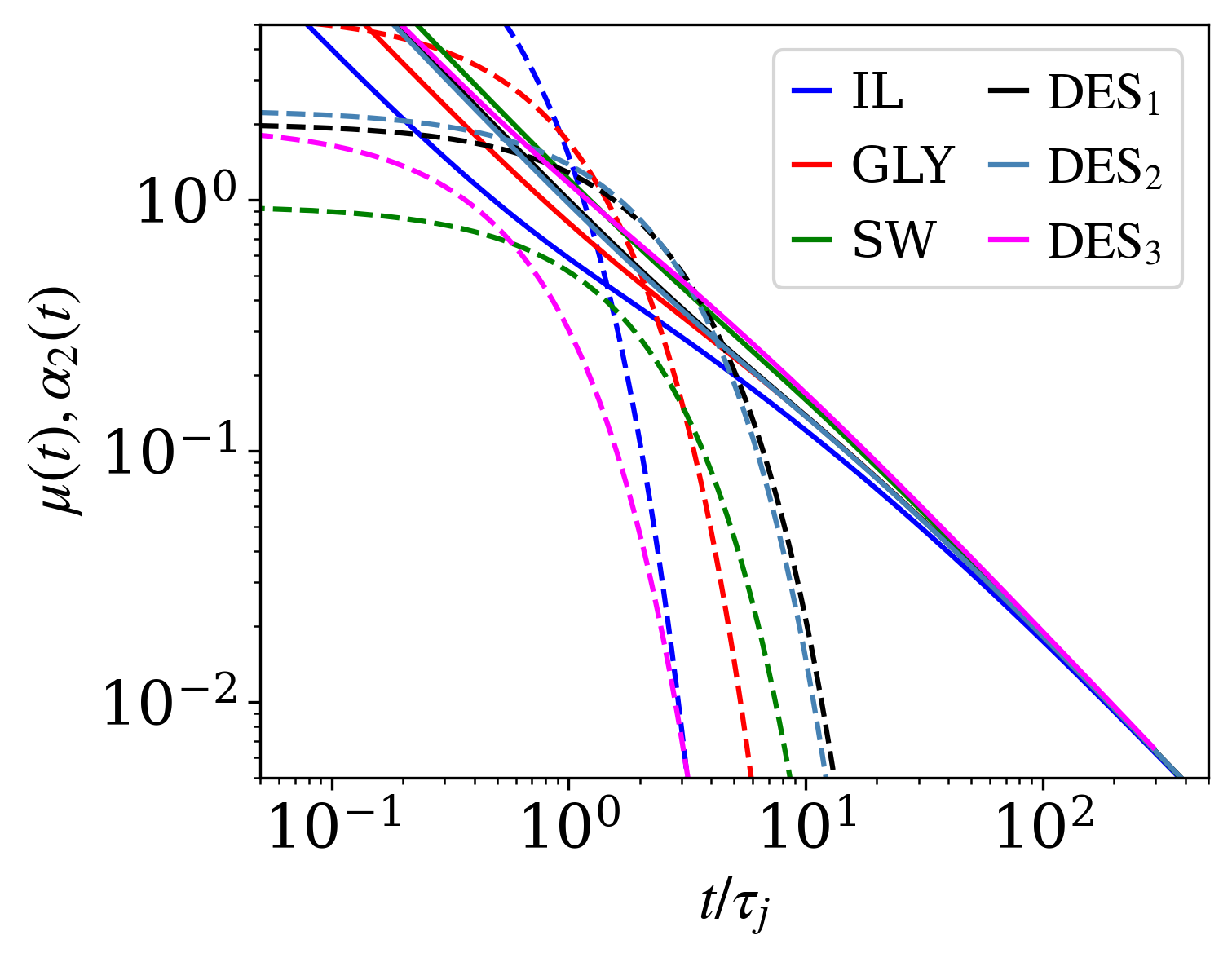}
    \caption{(top) Plots of NFP, $\mu(t)$ (dashed lines) and NGP $\alpha_2(t)$ (solid lines) with respect to $t/\tau_j$. These plots are obatined for different complex fluids from the parameters obtained through the IQENS spectra modelling.}
\end{figure}

\textbf{\textit{Non-Local Diffusion (NLD) model}}

We propose a general $d$-dimensional non-local diffusion (NLD) model to capture the FnGD regime. The Fokker-Planck equation for NLD is constructed by invoking non-local effects into the model. While Levy flights are a classical example of non-local processes, their scale-free property results in non-Gaussian behavior at all scales.

In our model, we introduce scaling parameters ($x_0, \tau_j$) that break the scale invariance, allowing the system to transition from non-Gaussian to Gaussian dynamics over sufficiently long distances and times. This approach enables the capture of both the non-Gaussian characteristics at short scales and the Gaussian behavior at macroscopic scales. The NLD equation governing such a process is given by,
\begin{equation}
\label{d-dimensional NLD}
\frac{\partial G_s(\mathbf{x},t)}{\partial t} = \frac{x_0^{2-d}}{\tau_j} \int d \mathbf{x}^\prime f\left[ \frac{|\mathbf{x}-\mathbf{x}^\prime|}{x_0}; t \right] \nabla_{\mathbf{x}'}^{2} G_s(\mathbf{x}^\prime, t)
\end{equation}
where $\mathbf{x} \in \mathbb{R}^d$ and $G_s(\mathbf{x},t)$ is the van Hove self-correlation function providing the probability associated to finding the particle at a position $\mathbf{x}$ at any given time $t$. The function, $f\big(|\mathbf{x}|/x_0; t\big)$ is a time-dependent jump kernel containing information about the non-local displacements, $x_0$ and $\tau_j$ are the characteristic length and time scales associated to the non-local diffusion process. With $f(|\mathbf{x}|) = x_0^d \delta(\mathbf{x})$ and $D=x_0^2/ \tau_j$, eq. \eqref{d-dimensional NLD} produces the standard $d$-dimensional Brownian motion. Meanwhile, power-law time-dependent kernel $f(|\mathbf{x}|,t) = t^{\alpha-1} f(|\mathbf{x}|)$ has been used to describe non-Gaussian fractional Brownian motion \cite{Srinivasan_2024}. In general, the solutions of eq. \eqref{d-dimensional NLD} can be realised in the form of SISF, 
\begin{equation}
\label{NLD FT soln}
I_s(Q, t) = I(Q,0) \exp\left[ - (Qx_0)^2 \hat{g}(Qx_0; t)\right]
\end{equation}
where $\hat g(Qx_0; t) = \tau_j^{-1} \int_0^t dt'\; \hat f(Qx_0; t')$ and $\hat f(Qx_0; t)$ is the radial Fourier transform of $f(|\mathbf{x}|/x_0)$. 

In order to solve eq. \eqref{d-dimensional NLD}, we recast this problem in the framework of a subordination scheme \cite{Chechkin_2021a, Gorenflo_2011a, Stanislavsky_2003a, Meerschaert_2002}. We consider the displacement of the particle obeying NLD to be $\mathbf{X}\left[\tau(t)\right]$, where  $\mathbf{X}(\tau)$ is a $d$-dimensional Wiener's process and $\tau(t)$ is a stochastic process with nonnegative increments. The SISF, $I_s(Q,t)$ is linked to the distribution of operational time $\tau(t)$, $T(\tau, t)$ through an integral decomposition formula\cite{Chechkin_2017, Chechkin_2021a, SM},
\begin{equation}
\label{subordination integral formula}
I_s(Q,t) = \int\displaylimits_0^\infty d\tau \: T(\tau, t) e^{-Q^2\tau} = \tilde T(u,t)\big|_{u=Q^2}
\end{equation}
where $\tilde T(u, t)$ is Laplace transform of $T(\tau, t)$. Upon comparing eqns. \eqref{NLD FT soln} and \eqref{subordination integral formula}, a one-one correspondence between $T(\tau, t)$ and $\hat g(Qx_0; t)$ is established through a inverse Laplace transform \cite{SM}: $T(\tau,t) = \mathcal{L}^{-1}\left[ \exp \left(- au \;\hat g(au;t)\right)\right]$, wherein $a \equiv x_0^2$. This relationship enables calculating $T(\tau, t)$ which when plugged back into the integral decomposition formula\cite{SM} directly provides $G_s(\mathbf{x},t)$.

Before, calculating the solutions to NLD, we study the necessary the conditions on jump-kernel for existence of FnGD. Considering the symmetry of the problem, the radial Fourier transform of the jump-kernel can in general be written as, $\hat f(Qx_0; t) = \sum_{n} c_n(t) (Qx_0)^{2n}$. For MSD to be linear in time, it follows that $c_0$ should be a non-zero real constant \cite{SM}. In addition, for the system to have non-Gausian behaviour at least $c_1(t)$ should be non-zero. However, for systems in equilibrium it should grow slower than $\mathcal{O}(t)$ to ensure that $\alpha_2(t)$ approaches zero asymptotically \cite{SM}. Combined these two requirement, form neccessary conditions for FnGD, although may not be sufficient. While a general case, encompasses time-dependent coefficients, a convergent series with time-independent coefficients $c_n$ can also produce FnGD characteristics \cite{Srinivasan_2024a}. In such cases, it can be shown that the $\alpha_2(t) \sim (\tau_j/t)$ \cite{SM}, which is akin to the asymptotic behaviour observed in the cage-jump diffusion in complex fluids. 

Therefore, in this letter, we consider a particularly relevant example, $c_n = (-1)^n$, which leads to the jump-kernel of the form $\hat f_e(Qx_0) = \left[1 + (Qx_0)^2\right]^{-1}$. Substituting this into eq. \eqref{NLD FT soln}, we can observe that it captures the behaviour of $I_{FnG}(Q,t)$ in eq. \eqref{iqt fngd}. Therefore, we study this particular kernel in greater detail using the subordination scheme. This kernel has exponential characteristic in the real domain \cite{SM, Srinivasan_2024a}. Particularly, for the 3D molecular jump-diffusion in complex fluids, explored in this letter, ($d=3$) $f_e(r/l_0) = (4\pi r)^{-1} l_0 e^{-r/l_0}$.
\begin{figure}
	\centering
	\includegraphics[width=\linewidth]{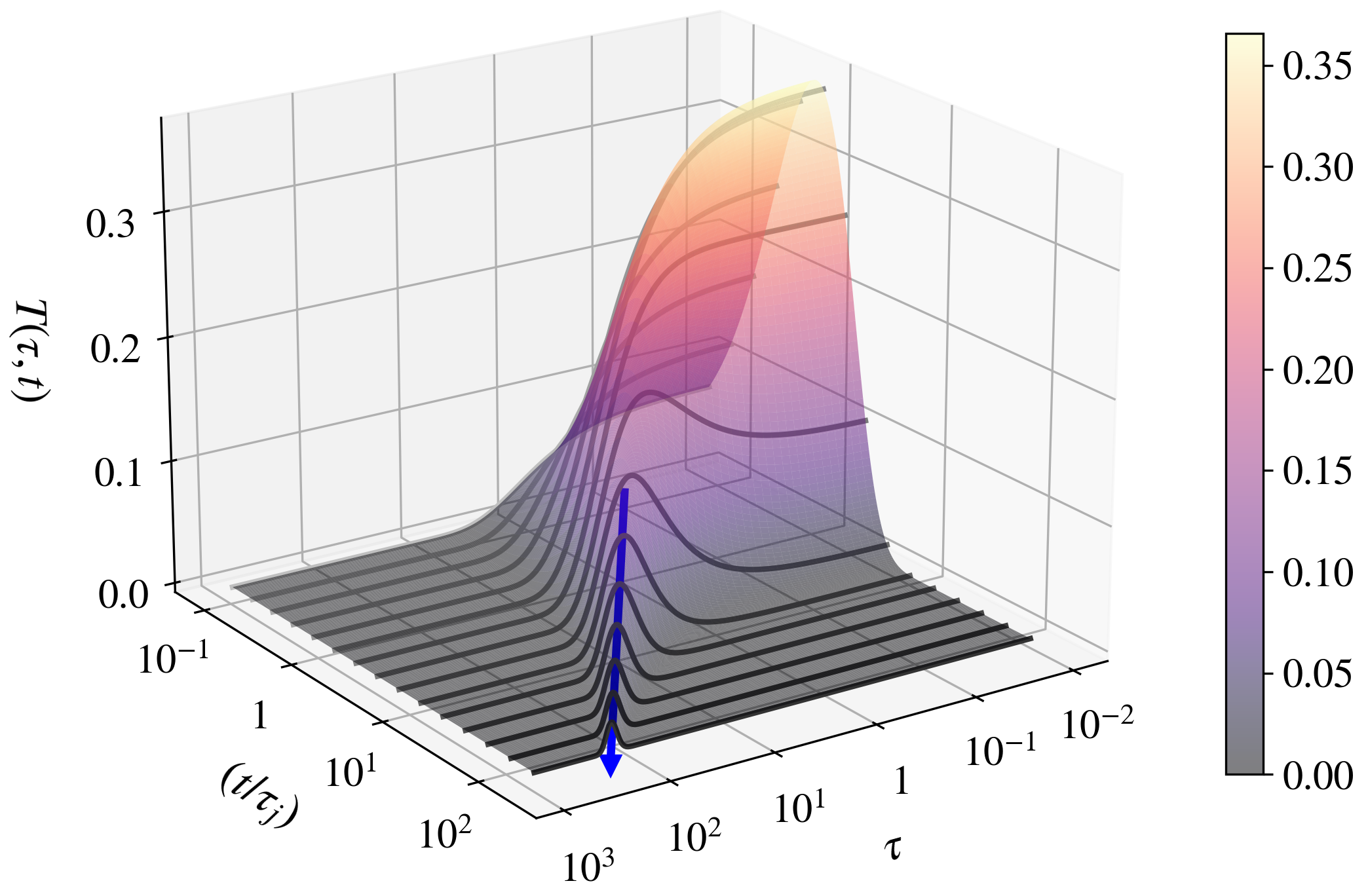}
	\caption{Evolution of the distribution associated to the subordinating process $\tau(t)$, $T(\tau, t)$ for the exponential kernel, $f_e(|\mathbf{x}|/x_0)$. The distribution evolves from being initially exponential in nature to a unimodal distribution which is centered around $\tau = D_j t$, which is indicated by the straight line (blue).}
	\label{tau distribution evolution}
\end{figure}
As this model effectively captures the physical jump-diffusion process observed in these complex fluids, we solve the $d$-dimensional NLD for $f_e(|\mathbf{x}|/x_0)$ using the proposed subordination scheme. 

The behaviour of jump-diffusion process can be captured through the calculation of  $T(\tau, t)$ which can be solved by inverting the Laplace transform, $T(\tau, t) = \mathcal{L}^{-1}\left\{ \exp\left[ - au/(1 + au) (t/\tau_j) \right] \right\}$. Therefore we get \cite{SM},
\begin{equation}
\label{exp kernel subordinator reduced}
\resizebox{0.9\hsize}{!}{$T(\tau,t) = e^{-\frac{t}{\tau_j}} \left[ \delta(\tau) + e^{-\frac{\tau}{x_0^2}} \sqrt{\frac{t}{x_0^2 \tau\tau_j}} I_1 \left(2 \sqrt{\frac{\tau}{x_0^2} \frac{t}{\tau_j}} \right)  \right]$}
\end{equation}
where $I_1(z)$ is the modified Bessel function of the first kind. An interesting pattern emerges when exploring the distribution of $T(\tau, t)$ at different values of $t/\tau_j$, as shown in Fig. 3. At timescales short compared to characterisitc jump time, $t \ll \tau_j$, we note that $T(\tau, t)$ typically exhibits an exponential behaviour, $e^{-\tau/x_0^2}$. Upon substitution into the integral decomposition formula this leads to exponential behaviour in the $G_s(\mathbf{x},t)$ \cite{SM, Chubynsky_2014}. This vindicates the emergence of exponential tails at timescales shorter or comparable to $\tau_j$. Meanwhile, for timescales large relative to characteristic jump time, $t \gg \tau_j$, $T(\tau, t)$ follows a unimodal distribution centered around $\tau = D_j t$. This is highlighted if Fig. 3, by tracking the peak of the distribution, with the line $\tau = D_j t$. Interestingly, in this regime, $T(\tau, t)$ has a width that varies $\sim (\tau_j/t)$ and therefore becomes narrower as $t/\tau_j \rightarrow \infty$, such that it achieves the limiting distribution $\delta\left(\sqrt{\tau} - \sqrt{D_j t}\right)$ \cite{SM}. Therefore, in this long-time, this kernel recovers the $d$-dimensional Brownian motion \cite{SM}.

Using eq. \eqref{exp kernel subordinator reduced} in the integral decomposition formula \cite{SM}, we calculate the radial van Hove self-correlation function, $g_s(r,t) = 2 \pi^{d/2} r^{d-1} G_s(\mathbf{x},t)/\Gamma(d/2)$, for a $d$-dimensional jump-diffusion process ($d \le 4$) \cite{SM},
\begin{equation}
\begin{gathered}
\label{gsrt d-dimensional soln}
g_s(r,t) = e^{-\frac{t}{\tau_j}} \Bigg[
 \delta(r) +
 \frac{4}{\Gamma(d/2)} 
\sum_{n=1}^\infty \frac{ (t/\tau_j)^n }{ n! (n-1)! } \\
\frac{1}{r}  \left( \frac{r} {2x_0} \right)^{n+\frac{d}{2}}
K_{n-\frac{d}{2}}\left( \frac{r}{x_0} \right) \Bigg]
\end{gathered}
\end{equation}
where $K_n(z)$ is the $n$-th order modified Bessel function of the second kind. In short time limit ($t \ll \tau_j$), considering $n=1$, we observe that $g_s(z,t) \sim z^{1-d/2}K_{1-d/2}(z)$. And in the asymptotic limit, $z = r/x_0 \rightarrow \infty$, we have exponential behaviour, $g_s(r,t) \rightarrow r^{(d-1)/2} e^{-r/x_0}$. This behaviour is characteristic of intermittent dynamics \cite{Barkai_2020} which is observed in jump-diffusion process, irrespective of the dimensionality the system. For $d=3$, which is relevant to the systems studied in this letter, using $K_{-1/2}(z) = \sqrt{\pi/2} e^{-z}/\sqrt{z}$ the exact short-time asymptotic follows,
\begin{equation}
g_s^{3d}(\mathbf{r}, t) = e^{-\frac{t}{\tau_j}}
\left[\frac{r\;e^{-r/r_0}}{4\pi x_0^2} \left[\frac{t}{\tau_j}\right]  +
\mathcal{O}\left( \left[\frac{t}{\tau_j}\right]^2 \right)  \right] \\
\end{equation}
entailing a linear increase near $r = 0$ and and exponential behaviour in large values of $r$. Our method ensues that follows from calculation of $T(\tau, t)$ enables the prediction of the asymptotic behaviour of $G_s(\mathbf{x},t)$ for any $d$-dimensional jump-diffusion process. 

\textbf{\textit{Conclusion}}
This letter presents a lucid study of FnGD in molecular diffusion in complex fluids using IQENS observations. The findings of this study pave way to understand the novel area of FnGD with a fresh perspective on the basis of cage-jump diffusion mechanism prevalent in these systems. The development of a $d$-dimensional NLD model and it's connection to the FnGD behaviour alternative and fresh perspective grounded on physical motivation of jump-diffusion process. Further, our work also highlights the effectiveness of employing subordination technique in solving NLD and therefore opens up avenue to solve numerous such jump-diffusion models through this connection. In particular, the connections between kernel of the NLD equation and the respective subordinating process for the diffusing-diffusivity model is an are of work that is pursued to expand the scope of these systems. Meanwhile, the diffusion in these complex fluids also exhibit a scale-dependent Stokes-Einstein breakdown \cite{Srinivasan_2024, Berrod_2017b}, whether these features emerge from the underlying FnGD nature of the system could be explored by studying more microscopic origin of the NLD.

\bibliography{MS_main}

\end{document}


\title{Supplementary Material for \\ Fickian Yet non-Gaussian Diffusion in Complex Molecular Fluids via a Non-Local Diffusion Framework}

\author{H. Srinivasan}
\affiliation{Solid State Physics Division, Bhabha Atomic Research Centre, Mumbai 400085, India}
\affiliation{Homi Bhabha National Institute, Anushaktinagar, Mumbai 400094, India}
 
\author{V. K. Sharma}
\affiliation{Solid State Physics Division, Bhabha Atomic Research Centre, Mumbai 400085, India}
\affiliation{Homi Bhabha National Institute, Anushaktinagar, Mumbai 400094, India}

\author{S. Mitra}
\affiliation{Solid State Physics Division, Bhabha Atomic Research Centre, Mumbai 400085, India}
\affiliation{Homi Bhabha National Institute, Anushaktinagar, Mumbai 400094, India}

\maketitle

\section{Overview of Incoherent Quasielastic Neutron Scattering (IQENS)}
Incoherent quasielastic neutron scattering (IQENS) technique captures information about the dynamics of the system. Traditionally, the description of IQENS spectra is conveniently expressed through the framework of van-Hove self-correlation functions \cite{Bee_1988}. The models of self-diffusion processes are also comprehensively captured in the description of van-Hove self-correlation function, $G_s(\mathbf{r}, t)$, which is defined as \cite{Bee_1988},
\begin{equation}
\label{van Hove defn}
G_s(\mathbf{r}, t) = \frac{1}{N} \left< \sum_{i=1}^N \delta \bigg[ \mathbf{r}  - \big( \mathbf{r}_i (t) - \mathbf{r}_i(0) \big) \bigg] \right>
\end{equation}
where $\mathbf{r}_i(t)$ is the position of the $i^{th}$ particle at any instant of time $t$ and $N$ is the total number of particles in the system. The angular brackets $\left< \right>$ indicate ensemble average. Intuitively, $G_s(\mathbf{r}, t)$ provides probability that the displacement of the particle is equal to $\mathbf{r}$ at any given time $t$, provided that the particle started at the origin at $t=0$.

The IQENS spectra measured in neutron scattering experiments is linked to the van-Hove self-correlation function through a double Fourier-transform in space and time,
\begin{equation}
S_{inc}(\mathbf{Q}, \omega) = \int\displaylimits_{-\infty}^{\infty}  dt \;e^{i\omega t} \int d\mathbf{Q} \;e^{-i\mathbf{Q}.\mathbf{r}} G_s(\mathbf{r},t) = \int\displaylimits_{-\infty}^{\infty} dt\;e^{i\omega t} I_s(\mathbf{Q}, t)
\end{equation}
where $\mathbf{Q}$ is the momentum-transfer and $\omega$ is the energy transfer in the neutron scattering experiment. Here $I_s(\mathbf{Q},t)$ is the self-intermediate scattering function (SISF) which is often convenient to describe the dynamics in complex fluids. The SISF, $I_s(\mathbf{Q},t)$, is defined according to,
\begin{equation}
\label{sisf defn}
I_s(\mathbf{Q},t) = \int d\mathbf{Q} \;
e^{-i\mathbf{Q}.\mathbf{r}} G_s(\mathbf{r},t)
\end{equation}
In is notable that, while working with liquids, the obtained spectra is considered as an isotropic average, essentially reducing to the IQENS spectra and SISF to be function of the magnitude of momentum transfer $|\mathbf{Q}| = Q$.

In the case of a particle undergoing Brownian motion, it is easy to show that $I_s(Q,t) = \exp\left[-D Q^2 t\right]$, wherein $D$ is the diffusivity of the particle. The IQENS spectra therefore, can be given by,
\begin{equation}
\label{sqw brownian}
S_{inc}(Q, \omega) = \frac{1}{\pi}\frac{DQ^2}{\left[DQ^2 \right]^2 + \omega^2}
\end{equation}
which essentially indicates the profile of peak will correspond to a Lorentzian, whose half-width half-maximum, $\Gamma(Q)$ varies as $DQ^2$ with respect to $Q$. These are the central features observed in the IQENS spectra for Brownian motion.

Before, proceeding further, it is important to highlight the behaviour of SISF and IQENS spectra, for the case of particles simultaneously executing two independent dynamical degrees of freedom. Typically, these correspond to localized and long-range diffusion processes, with the associated SISFs $I_{loc}(Q,t)$ and $I_{long}(Q,t)$. The effective SISF can be written as a product of individual components,
\begin{equation}
\label{combination iqt}
I_s(Q,t) = I_{loc}(Q,t) I_{long}(Q,t)
\end{equation}
This is equivalent to convolution in the energy space, indicating that $S_{inc}(Q, \omega) = S_{loc}(Q, \omega) \otimes S_{long}(Q, \omega)$.

\newpage
\section{Cage-Jump diffusion model}
The cage-jump diffusion model is a composition of two dynamical modes involving jump and caged diffusion processes \cite{Qvist_2011, Embs_2012, Burankova_2014, Srinivasan_2020, Wagle_2015}. After obtaining the SISFs associated to the jump and caged diffusion models, we can multiply them to obtain the SISF for cage-jump diffusion model owing to the independence of their degrees of freedom.

\subsection{Jump diffusion model}
The jump diffusion process can be explicitly described through the SISF of the form \cite{Qvist_2011, Srinivasan_2020}
\begin{equation}\label{jump}
I_j(Q,t) = \exp\left[ \frac{-(Q l_0)^2}{1 + (Q l_0)^2} \frac{t}{\tau_j} \right]
\end{equation}
where $l_0$ is the characteristic jump length and $\tau_j$ is the average time interval between jumps. It is evident that the IQENS spectra in this case will correspond to a Lorentzian,
\begin{equation}
S_j(Q, \omega) = \frac{1}{\pi}\frac{\Gamma_j(Q)}{\Gamma_j(Q)^2 + \omega^2}
\end{equation}
where $\Gamma_j$ is the HWHM of Lorentzian. The HWHM follows the relationship
\begin{equation}
\Gamma_j(Q) = \frac{1}{\tau_j}\frac{ (Ql_0)^2}{1 + (Ql_0)^2} = \frac{D_j Q^2}{1 + \tau_j D_j Q^2}
\end{equation}
where $D_j = (l_0^2/\tau_j)$ is the jump diffusivity of the particle.

\subsection{Caged diffusion model}
The caged diffusion process can be described through localized translational diffusion within an isotropic gaussian well \cite{Burankova_2014, Volino_2006}. This model physically captures the dynamics of a particle in soft containment and provides analytically smooth solutions across all space, unlike hard confinement potentials derived earlier \cite{Srinivasan_2020, Volino_1980}. The SISF of this process is given as \cite{Volino_1980},
\begin{equation}\label{localized_raw}
I_{loc}(Q,t) = \exp \left[ -(Q \sigma)^2 \left(1-e^{-t/\tau_0} \right) \right]
\end{equation}
where $\sigma$ is the characteristic localization radius and $\tau_0$ is the relaxation timescale associated to localized diffusion process. Given the heterogeneity arising due to structural heterogeneity in the molecular system, one can expect a variation in the radius of localization. We incorporate this variation by considering a probability distribution $p(\sigma; \sigma_0)$, with a mean value $\sigma_0$ . In such a scenario, the SISF is given as,
\begin{align}\label{localized}
I_{loc}(Q,t) = \int_0^\infty d\sigma \;p(\sigma; \sigma_0) \exp \left[ -(Q \sigma)^2 \left(1- e^{-t/\tau}\right) \right] \\
\label{localized_avg}
= \left< \exp \left[ -(Q \sigma)^2 \left(1- e^{-t/\tau}\right) \right] \right>_\sigma
\end{align}
where $\left<\right>_\sigma$ denotes average over the distribution of $\sigma$.
The above equation can be expanded in series to obtain the time-dependent and independent parts, 
\begin{align}\label{loc_expansion}
\begin{aligned}
I_{loc}(Q,t) = \int_0^\infty d\sigma \;p(\sigma; \sigma_0) e ^{-(Q\sigma)^2} e^{(Q\sigma)^2 e^{-t/\tau_0}} \\
	= \int_0^\infty d\sigma \;p(\sigma; \sigma_0) e ^{-(Q\sigma)^2} \sum_{n=0}^\infty \frac{(Q\sigma)^{2n}}{n!} (e^{-t/\tau_0})^n  \\
	= A_0(Q) + \sum_{n=1}^\infty A_n(Q) e^{-(nt/\tau_0)}
 \end{aligned} 
\end{align}
where $A_0(Q)$ and $A_n(Q)$ are the elastic incoherent structure factor (EISF) and quasielstic incoherent structure factor respectively, defined as,
\begin{align}\label{eisf}
A_0(Q) = \int_0^\infty d\sigma \;p(\sigma; \sigma_0) e^{-(Q\sigma)^2} = \left<e^{-(Q\sigma)^2}\right>_\sigma \\
\label{qisf}
A_n(Q)	= \int_0^\infty d\sigma \;p(\sigma; \sigma_0) e ^{-(Q\sigma)^2} \frac{(Q\sigma)^{2n}}{n!} = \frac{\left< e ^{-(Q\sigma)^2}(Q\sigma)^{2n}  \right>_\sigma}{n!}
\end{align}
\subsection{Composing jump and caged diffusion processes}
In the systems under consideration, the caged-diffusion process is happening continuously with sudden jumps in it's position from time to time, dictacted through the mean waiting times $\tau_j$. Noting that the localized process and the jump process are independent, the effective SISF for the cage-jump diffusion process is given by the product of their individual SISF. 
Combining eq \eqref{jump} and \eqref{localized_avg}, we have the SISF for the cage-jump diffusion model to be given by,
\begin{equation}\label{complete}
I_{cj}(Q,t) = I_j(Q,t) I_{loc}(Q,t) = \exp\left[ \frac{-(Q l_0)^2}{1 + (Q l_0)^2} \frac{t}{\tau_j} \right]\left< \exp \left[ -(Q \sigma)^2 \left(1- e^{-t/\tau_0} \right) \right] \right>_\sigma 
\end{equation}
In subsequent sections we will derive QENS spectra related to cage-jump model using the Fourier transform of \eqref{complete}.

\subsection{Derivation of MSD, NFP and NGP}

Given that we have obtained SISF for cage-jump diffusion mechanism, $I_{cj}(Q,t)$, it is possible to calculate the moments of the distribution from which we can obtain the expressions for mean-squared displacement (MSD) $\left< \delta r^2(t) \right>$ and the non-Gaussian parameter (NGP) $\alpha_2(t)$ for cage-jump model. The moments of displacements are related to $I(Q,t)$ as,
\begin{equation}\label{moments}
\left< \delta r^{2n} (t) \right> = (-1)^n(\nabla_Q)^{2n} I(Q,t) |_{Q=0}
\end{equation}
Using this formula in eq \eqref{complete}, we get the MSD as,
\begin{equation}\label{msd_complete}
\left< \delta r^2 (t) \right> = 6 \left[ \left(1- e^{-t/\tau_0}\right) \left< \sigma^2 \right>_\sigma + D_j t \right]
\end{equation}
where we have written $D_j = \frac{l_0^2}{\tau_j}$, as the jump diffusion constant. In order to characeterize the extent of Fickianity, we define a parameter non-Fickian parameter (NFP), $\mu(t)$ as,
\begin{align}
\label{nfp}
\begin{aligned}
\mu(t) = \frac{1}{6D_j}\frac{d\left< \delta r^2 (t) \right> }{dt} - 1 \\
  = \frac{1}{D_j}\bigg[\frac{\left< \sigma^2 \right>_\sigma }{\tau_0} e^{-t/\tau_0} + D_j \bigg] - 1
\end{aligned}
\end{align}
such that, $\mu(t) = 0$ for Fickian diffusion process. It is notable that the restoration to Fickianity in the system is particularly dependent on the relaxation timescale of localized diffusion process governed by $\rho(t)$.

In a similar fashion, we can obtain the NGP, $\alpha_2(t)$ which characterizes the deviation from Gaussian nature of the displacement distribution,
\begin{align}\label{ngp_complete}
\begin{aligned}
\alpha_2(t) = \frac{3}{5}\frac{\left< \delta r^4(t) \right>}{\left[\left< \delta r^2 (t) \right>\right]^2} - 1 = \frac{2 \tau_j}{t} \left< \frac{1}{\left[1 + \frac{\sigma^2}{l_0^2} \frac{\tau_j}{t} \left[1- e^{-t/\tau_0} \right]\right]^2} \right>_\sigma \\
= 2 \bigg(\frac{\tau_j}{t} \bigg) \sum_{n=0}^\infty (-1)^n \frac{\left< \sigma^{2n}\right>_\sigma}{l_0^{2n}}\left( \frac{\tau_j}{t} \right)^n \left[ 1- e^{-t/\tau_0} \right]^{2n}
\end{aligned}
\end{align}
It is notable that the restoration of Gaussianity is dictated by the slowest term in the above equation which decays $(\tau_j/t)$. Therefore, the onset of Gaussian regime the system's dynamics is substantially controlled by, $\tau_j$, mean waiting time between the jumps.

\subsection{IQENS model for cage-jump diffusion}
In order to obtain incoherent quasielastic neutron scattering (IQENS) spectra, we combine eq \eqref{jump} and \eqref{loc_expansion},
\begin{equation}\label{complete_iqt}
I(Q,t) = \exp\left[ \frac{-(Q l_0)^2}{1 + (Q l_0)^2} \frac{t}{\tau_j} \right] \left[A_0(Q) + \sum_{n=1}^\infty A_n(Q) e^{-nt/\tau_0} \right]
\end{equation}
Transforming \eqref{complete_iqt} to energy domain, we obtain the incoherent quasielastic neutron scattering (IQENS) spectra for the cage-jump diffusion model,
\begin{align}
\label{complete_sqe}
\begin{aligned}
S(Q,\omega) = L\left(\Gamma_j, \omega \right) \otimes \left( A_0(Q) \delta(\omega) + \sum_{n=0}^\infty A_n(Q) L\left(\Gamma_{n}, \omega \right) \right) \\
=  A_0(Q) L\left(\Gamma_j, \omega \right)  + \sum_{n=0}^\infty A_n(Q) L\left(\Gamma_{n} + \Gamma_{j}, \omega \right) 
\end{aligned}
\end{align}
where $L\left(\Gamma, \omega \right)$ is a Lorentzian with half-width half-maximum (HWHM) of $\Gamma$. The parameters $\Gamma_j(Q) = \frac{D_j Q^2}{1+\tau_j D_j Q^2}$ and $\Gamma_n(Q) = n/\tau_0$ are the HWHM associated to jump and caged processes.

While, eq. \eqref{complete_sqe} gives the complete equation for a cage-jump diffusion model. A simpler two-component model function is typically used in the fitting IQENS data is given by \cite{Qvist_2011, Srinivasan_2020, Srinivasan_2023, Srinivasan_2024a}, 
\begin{equation}
\label{iqens_model}
S_{mod}(Q,\omega) = C(Q) L\left(\Gamma_j, \omega \right) + \left[1-C(Q)\right] L\left(\Gamma_{loc}+\Gamma_j, \omega \right) 
\end{equation}
The robustness of this model has also been verified in a number of different systems including supercooled water \cite{Qvist_2011}, glyceline DES \cite{Wagle_2015}, ionic liquids \cite{Embs_2012, Burankova_2014, Burankova_2017} and deep eutectic solvents \cite{Srinivasan_2020, Srinivasan_2023}. Comparing \eqref{complete_sqe} and \eqref{iqens_model}, it is immediately evident that $C(Q)$ is the EISF of the localized diffusion process $A_0(Q)$. Secondly, it is notable that the series of Lorentzians in \eqref{complete_sqe} is approximated by a single Lorentzian of HWHM $\Gamma_{loc} + \Gamma_j$. While an analytical expression for $\Gamma_{loc}(Q)$ cannot be obtained, we later evaluate it numerically based on the complete series given in eq. \eqref{complete_sqe} to estimate the parameter $\tau_0$ from the fits.

Using an exponential distribution $p(\sigma;\sigma_0) = \sigma_0^{-1} e^{-\sigma/\sigma_0}$ for $\sigma$, we further evaluate the terms $A_0(k)$ and $A_n(k)$ in \eqref{complete_sqe}. The EISF in this case is given by,
\begin{align}
\begin{aligned}
A_0(Q) = \int_0^\infty d\sigma \frac{e^{-\sigma/\sigma_0}}{\sigma_0} e^{-(Q\sigma)^2} \\
 = \frac{\exp \left[\frac{1}{4Q^2\sigma_0^2}\right] \sqrt{\pi}\mathrm{Erfc}\left[\frac{1}{2Q\sigma_0}\right]}{2Q\sigma_0}
\end{aligned}
\end{align}
where $\mathrm{Erfc}[z]$ is the complementary error function. We use this model to fit the EISF obtained from IQENS experiments with $\sigma_0$ as the fitting parameter. We find all the systems (including the DESs) are very well described in this model.

\begin{figure}
	\centering
	\includegraphics[width=0.5\linewidth]{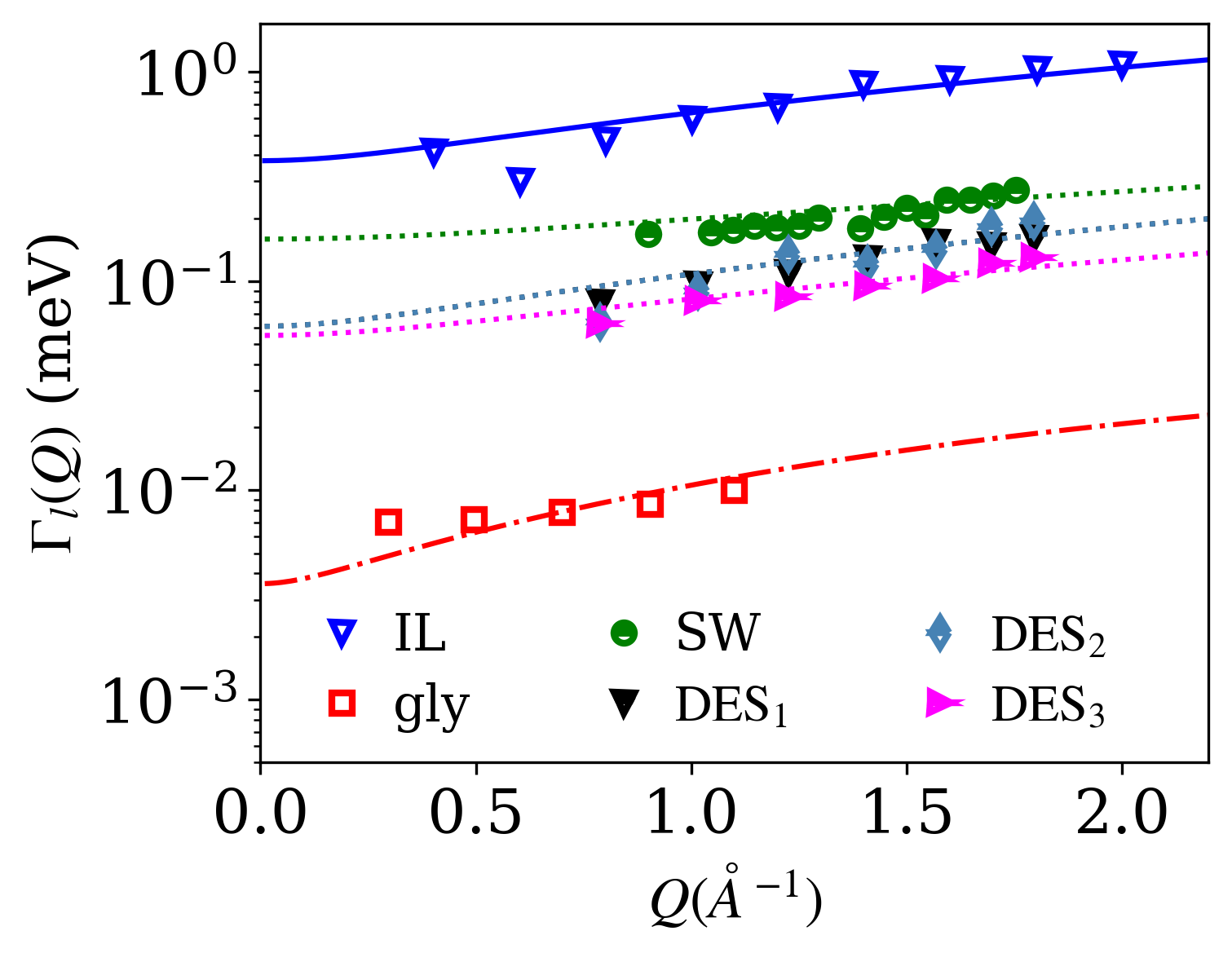}
	\caption{(top) The variation of relaxation rate associated to the localized diffusion $\Gamma_{loc}(Q)$, with respect to $Q$ for different systems surveyed in this study (IL: ionic liquid; gly: glycerol; SW: supercooled water; $\text{DES}_i$: deep eutectic solvents. The fits based on diffusion within soft sphere for localized process is also shown along side IQENS data points.}
\end{figure}

\subsection{Approach to Fickian yet non-Gaussian diffusion (FnGD)}
The approach to Fickian yet non-Gaussian diffusion (FnGD) regime can be characterized through the non-Fickian parameter (NFP) is given by,
\begin{equation}
\label{nfp_fngd}
\mu(t) = \frac{\left< \sigma^2 \right>_\sigma }{l_0^2} \frac{\tau_j}{\tau_0} e^{-t/\tau_0}
\end{equation}
It is clear that for $t \gg \tau_0$, NFP goes to zero at an exponential rate, governed by the timescale associated to localized diffusion $\tau_0$. Therefore, we can define $t \gg \tau_0$ as the limit to approach Fickian diffusion. On the other hand the non-Gaussian parameter (NGP) is,
\begin{equation}
\label{ngp_complete_e}
\alpha_2(t) = 2\frac{\tau_j}{t} \left< \left[1 + \frac{\sigma^2}{l_0^2} \frac{\tau_j}{t}   \left(1-e^{-t/\tau_0}\right) \right]^{-2} \right>_\sigma
\end{equation}
which decays at a much slower rate governed by $\tau_j$ and therefore it remains non-zero even in the Fickian diffusion limit ($t \gg \tau_0$). In this limit, it is clear the MSD is linear and the NGP is given by,
\begin{equation}
\label{ngp_fngd}
\alpha_2(t) = 2\frac{\tau_j}{t} \left< \left[1 + \frac{\sigma^2}{l_0^2} \frac{\tau_j}{t}\right]^{-2} \right>_\sigma
\end{equation}
clearly indicating non-zero NGP with $1/t$ decay. For sufficiently long times $t \gg \tau_j$, we can further expand the above equation binomially giving us
\begin{align}
\label{ngp_fngd_long}
\alpha_2(t) \approx 2\frac{\tau_j}{t}  \left[1 - 2 \frac{\left<\sigma^2\right>_\sigma}{l_0^2} \frac{\tau_j}{t}\right]
\end{align}
Substituting the limit $t \gg \tau_0$ in eq. \eqref{complete_iqt}, we get the SISF for the Fickian yet non-Gaussian (FnGD) limit,
\begin{align}\label{iqt_fngd}
\begin{aligned}
I_{FnG}(Q,t) = \exp\left[ \frac{-(Q l_0)^2}{1 + (Q l_0)^2} \frac{t}{\tau_j} \right]\left< \exp \left[ -(Q \sigma)^2 \right] \right>_\sigma \\
= A_0(Q) \exp\left[ \frac{-(Q l_0)^2}{1 + (Q l_0)^2} \frac{t}{\tau_j} \right]
\end{aligned}
\end{align}
where we have used eq \eqref{eisf} to obtain it in the form of EISF. We will show in the next section that our non-local jump diffusion gives solution of the form given in eq. \eqref{iqt_fngd}, when an exponential jump kernel is chosen.

\newpage
\section{Non-local diffusion (NLD) model}
We introduce the $d$-dimensional non-local diffusion (NLD) model through the Fokker-Planck equation,
\begin{equation}
\label{d-dimensional NLD}
\frac{\partial G_s(\mathbf{x},t)}{\partial t} = \frac{x_0^{2-d}}{\tau_j} \int d \mathbf{x}^\prime \; f\left( \frac{|\mathbf{x}-\mathbf{x}^\prime|}{x_0}; t \right) \nabla^{'2} G_s(\mathbf{x}^\prime, t)
\end{equation}
where $\mathbf{x} \in \mathbb{R}^d$ is a vector in the $d$-dimensional space and $G_s(\mathbf{x},t)$ is the van Hove self-correlation function providing the probability associated to finding the particle at a position $\mathbf{x}$ at any given time $t$. The function, $f\big(|\mathbf{x}|/x_0; t\big)$ is a time-dependent jump kernel containing information about the non-local displacements, $x_0$ and $\tau_j$ are the length and time scales associated to the non-local jump like diffusion process. The choice considering the kernel to have only radial dependence, $|\mathbf{x}|$, reflects the isotropic nature of diffusion process. In order to obtain the solutions to this equation in the Fourier space, we define the $d$-dimensional Fourier and inverse-Fourier transforms according to
\begin{align}
\label{d-dimensional FT}
\begin{aligned}
\hat\phi(\mathbf{Q}) = \int d \mathbf{x}\; e^{-i \mathbf{Q}.\mathbf{x}} \phi(\mathbf{x}) \\
\phi(\mathbf{x}) = \frac{1}{(2\pi)^d} \int d \mathbf{Q} \; e^{i \mathbf{Q}.\mathbf{x}} \phi(\mathbf{Q})
\end{aligned}
\end{align}

 wherein the SISF is defined according to,
\begin{equation}
\label{iqt FT defn}
I_s(\mathbf{Q},t) = \int d\mathbf{x} \; e^{-i \mathbf{Q}.\mathbf{x}} G_s(\mathbf{x},t)
\end{equation}

The eq. \eqref{d-dimensional NLD} are represented easily in the Fourier domain,
\begin{align}
\label{fpe_ft_fngd}
\begin{aligned}
\frac{\partial I_s(\mathbf{Q},t)}{\partial t} = - \frac{(Qx_0)^2}{\tau_j} \hat{f}(Qx_0; t)I(\mathbf{Q},t)
\end{aligned}
\end{align}
where $\hat f(Q x_0) = x_0^{-d} \hat{f}(Q; t)$ (here $\hat{f}(Q;t)$ is the Fourier transform of the jump-kernel $f(|\mathbf{x}|;t)$ defined according to eq. \eqref{d-dimensional FT}). It is important to note that the Fourier transform of the jump-kernel is $\hat{f}(Q)$ is essentially a radial function in $Q \equiv |\mathbf{Q}|$, which emerges due to the radial property of the jump-kernel itself.

Integrating the above equation, we have the general solution to the $d$-dimensional NLD,
\begin{equation}
\label{NLD FT soln}
I_s(\mathbf{Q}, t) = I_0(\mathbf{Q}) \exp\left[ - (Qx_0)^2 \hat{g}(Qx_0; t)\right]
\end{equation}
where $I_0(\mathbf{Q})$ is the Fourier transform of the initial distribution before any random jump and $\hat{g}(Qx_0;t)$ is defined according to,
\begin{equation}
\hat{g}(Qx_0; t) = \frac{1}{\tau_j}\int\displaylimits_{0}^{t} \hat f(Qx_0; t) dt
\end{equation} 
Without loss of generality, we can consider an initial condition, $G_s(\mathbf{r},0) = \delta(\mathbf{r})$, in which case the eq. \eqref{NLD FT soln} becomes,
\begin{equation}
\label{NLD FT soln 2}
I_s(Q, t) = \exp\left[ - (Qx_0)^2 \hat{g}(Qx_0; t)\right]
\end{equation}
Yet again, we note that the SISF is now only a function of the radial component $|\mathbf{Q}| \equiv Q$, which is due to the isotropic nature of the diffusion process.

\subsection{The case of Brownian motion}
In the case of infinitesimally small displacements, we consider the delta function kernel, which is $f(\mathbf{x} - \mathbf{x}') = \delta(\mathbf{x} - \mathbf{x}')$, and therefore eq. \eqref{d-dimensional NLD} reduces to the usual diffusion equation and the SISF has the form, $I(Q,t) = e^{-D_j Q^2 t}$, using $D_j = (x_0^2/\tau_j)$. Upon inverse Fourier transform, this leads to the $d$-dimensional Brownian motion model,
\begin{equation}
G_s(\mathbf{x},t) = \frac{1}{(4\pi D_j t)^{d/2}} \exp\left[ \frac{-\mathbf{x}^2}{4 D_j t}\right]
\end{equation}

\newpage
\section{Solutions through subordination}
Models of subordinated stochastic processes have been observed to be useful in systems exhibiting a distribution relaxation timescales \cite{Chechkin_2017, Chechkin_2021, Meroz_2015, Gorenflo_2011}, particularly pertaining to systems with intrinsic heterogeneity in the environment of the diffusion process. We employ the technique of subordination here to obtain greater insights about the nature of solutions from the NLD equation. 

Consider the particle following mechanism of diffusion described through NLD (eq. \eqref{d-dimensional NLD}), be described through a stochastic process $\mathbf{X}(t)$. We consider, $\mathbf{X}\left[\tau(t)\right]$ to be subordinated by another stochastic process $\tau(t)$ in such a way that $\mathbf{X}(\tau)$ is a Wiener's process. The combined set of Langevin equations, describing these two processes are given by \cite{Chechkin_2017},

\begin{align*}
\frac{d\mathbf{X}(\tau)}{d\tau} = \sqrt{2} \boldsymbol{\eta} (\tau) \\
\frac{d\tau(t)}{dt} = \xi (t) 
\end{align*}
where $\boldsymbol{\eta}(\tau)$ is the $d$-dimensional Gaussian white noise process, whose autocorrelation is given by $\langle\eta_i(\tau) \eta_j(\tau') \rangle = \delta_{ij}\delta(\tau - \tau')$ and $\xi(t)$ is a non-negative stochastic process guarenteeing non-negative random increments of $\tau(t)$. As $X(\tau)$ is a Wiener's process in $d$-dimensional space, it's solutions are given as,
\begin{equation}
P(\mathbf{x}, \tau) = \frac{1}{(4 \pi \tau)^{d/2}} \exp\left[-\frac{\mathbf{x}^2}{4\tau}\right]
\end{equation}
where $\mathbf{x}^2 = |\mathbf{x}|^2 = \sum_i^d x_i^2$ is the Euclidean distance in $\Bbb R^d$ space. Meanwhile, using the integral transformation for the subordination formula, we can obtain the PDF of $X(t)$ as,
\begin{equation}
\label{subordination integral transform}
G_s(\mathbf{x},t) = \int\displaylimits_{0}^\infty d\tau\; T(\tau, t) P(\mathbf{x},\tau)
\end{equation}
Upon Fourier transform of the above equation, we obtain the relationship,
\begin{equation}
I_s(Q,t) = \int\displaylimits_{0}^\infty d\tau\; T(\tau, t) e^{-Q^2 \tau}
\end{equation}
Noting that the above equation is akin to Laplace transform in for $Q^2 \equiv u$, we rewrite the above equation as,
\begin{equation}
\label{subordination integral}
I_s(Q,t) = \int\displaylimits_{0}^\infty d\tau\; T(\tau, t) e^{- u \tau} \bigg|_{u=Q^2} = \mathcal{L}_{\tau \rightarrow u} \left\{ T(\tau, t) \right\} = \tilde{T}(u, t)\bigg|_{u=Q^2}
\end{equation}
Therefore, we note that the distribution of $\tau$ can be written as inverse Laplace transform of $I(\mathbf{Q},t)$ for $Q = \sqrt{u}$, 
\begin{equation}
T(\tau, t) = \mathcal{L}^{-1} \left\{ I(Q = \sqrt{u}, t) \right\}
\end{equation}
Comparing the above equation with the solutions of $d$-dimensional NLD in the Fourier space (eq. \eqref{NLD FT soln} ), we can write
\begin{equation}
\label{subordination connection}
T(\tau, t) = \mathcal{L}^{-1}\left\{ \exp\left[-a u\; \hat{g}(au;t)\right]\right\}
\end{equation}
where we have used $Q^2 = u$ and $x_0^2 = a$ for convenience of calculations. \textbf{It is important to note that the above equation presents one-one correspondence between the distribution of subordinating parameter time, $T(\tau, t)$ and the jump-kernel of the NLD equation $\hat{g}(Qx_0; t)$.} This will serve as an important tool in trying obtain solutions for different jump-kernels later.

\vspace{0.5cm}
The case of Brownian motion can be easily recovered by considering $\hat{f}(Q x_0) = 1$ and therefore $\hat{g}(Qx_0) = (t/\tau_j)$. Therefore, we have $T(\tau, t)$ given by,
\begin{equation}
T_B(\tau,t) = \mathcal{L}^{-1}\left\{ exp\left[ -a u (t/\tau_j) \right]\right\} = \delta(\tau - a t/\tau_j) = \delta(\tau - D_j t)
\end{equation}
where we have used $a/\tau_j = x_0^2/\tau_j \equiv D_j$. Using this in eq. \eqref{subordination integral}, we have the solution for $\mathbf{X}(t)$,
\begin{equation}
G_s(x,t) = \int\displaylimits_{0}^\infty d\tau \; \delta(\tau - D_j t) \frac{1}{(4\pi\tau)^{d/2}} \exp\left[ \frac{-\mathbf{x}^2}{4\tau}\right] = \frac{1}{(4\pi D_j t)^{d/2}} \exp\left[ \frac{-\mathbf{x}^2}{4 D_j t}\right]
\end{equation}

\subsection{Conditions for FnGD}
While, a general NLD model can provide a variety of diffusion mechanisms depending on the choice the kernel, here we particularly focus on the nature of kernel that can lead to the behaviour observed in FnGD processes. Essential features of this include the following points
\begin{itemize}
	\item The mean-squared displacement (MSD) of the particle is always linear in time, $\langle \mathbf{x}^2(t) \rangle \sim t$.
	\item The non-Gaussianity parameter, $\alpha_2(t)$, is non-zero despite linear MSD
	\item While the displacement distribution is non-Gaussian at short times compared to jump-dynamics ($t \ll \tau_j$), the system eventually goes into the Gaussian limit at long-times ($t \gg \tau_j$).
\end{itemize}

While the last condition is not absolutely necessary for FnGD, it becomes an important criteria for most physical systems in equilibrium, as the systems tend to Gaussian behaviour at long times.

In the following, we show the necessary mathematical conditions required to satisfy the listed requirements by considering a power-series expansion of the radial transform of the jump-kernel, $\hat{f}(Qx_0;t)$. Since we are considering isotropic processes, we can write in general $\hat{f}(Qx_0;t)$ as even function of $(Qx_0)^2$. In general
\begin{equation}
\label{kernel_gen}
\hat f(Qx_0; t) = \sum_{n=0}^\infty c_n(t) (Qx_0)^{2n}
\end{equation}
where $c_n(t)$ are arbitrary coefficients which time-dependent. Substituting the above general kernel in eq. \eqref{NLD FT soln 2}, we have the SISF,
\begin{equation}
\label{general iqt}
I_s(Q,t) = \exp \left[ -\sum_{n=0}^\infty b_n(t) (Qx_0)^{2(n+1)} \right]
\end{equation}
where coefficients, $b_n(t) = \int_0^t dt' c_n(t')$, which appear in the corresponding integrated jump-kernel $\hat{g}(Qx_0; t)$.

Using the relationship shown in eq. \eqref{subordination integral}, we can rewrite the above equation according to,

\begin{equation}
\label{general LT subordinator}
\tilde{T}(u,t) = \exp\left[ - \sum_{n=0}^\infty b_n(t) (a u)^{n+1} \right]
\end{equation}
which is the Laplace transform of the distribution associated to the subordinating stochastic process, $\tau(t)$. Written in the above form, it is convenient to calculate the mean-squared displacement and non-Gaussian parameters using the formulae listed below \cite{Chechkin_2017},
\begin{equation}
\begin{gathered}
\langle \mathbf{x}^2(t) \rangle = -2d \frac{\partial \tilde{T}(u,t)}{\partial u}\bigg|_{u=0} \\
\langle \mathbf{x}^4(t) \rangle = 4d(d+2) \frac{\partial^2 \tilde{T}(u,t)}{\partial u^2}\bigg|_{u=0} 
\end{gathered}
\end{equation}

Using the above formulae and calculating for the general kernel using eq. \eqref{general LT subordinator}, we have,
\begin{equation}
\begin{gathered}
\langle \mathbf{x}^2(t) \rangle = 2d\; b_0(t) a \\
\langle \mathbf{x}^4(t) \rangle = 4d(d+2) \left[ b_0(t)^2 - 2b_1(t) \right]a^2 \\
\end{gathered}
\end{equation}
It is clear from the above equation, for MSD to have linear time-dependence, $b_0(t) \sim t$, indicating that $c_0(t)$ should be a non-zero real constant, i.e. $c_0(t) \rightarrow c_0$. Further, to calculate the non-Gaussian parameter, we use the general formula for the $d$-dimensional system,
\begin{align}
\alpha_2(t) = \frac{d}{d+2} \frac{\langle \mathbf{x}^4(t)}{(\mathbf{x}^2(t))^2} - 1 \implies
\alpha_2(t) &= \frac{-2b_1(t)}{b_0(t)^2} 
\end{align}
indicating that $b_1(t) < 0$ for obtaining a positive non-Gaussian parameter, which typically occurs in physical systems involving large-jumps. Further, it has to be ensured that $b_1(t)$ doesn't grow faster than $\mathcal{O}(t^2)$, since in that case the non-Gaussian parameter will increase with time, which seems unphysical for a system in thermal equilibrium. While, $b_1(t) \sim t^2$ will essentially correspond to a constant value of $\alpha_2$.

Further, setting $c_0 \ne 0$ in the series expansion, eq \eqref{kernel_gen}, we observe that in the long-wavelenght limit ($Qx_0 \rightarrow 0$)
\begin{equation}
\lim_{Qx_0 \rightarrow 0} \hat f(Qx_0) = c_0
\end{equation}
which can be rewritten in the real-space as,
\begin{equation}
\lim_{|\mathbf{x}|/x_0 \rightarrow \infty} f(\mathbf{x})= c_0 \delta(\mathbf{x})
\end{equation}
which corresponds to kernel associated to the Brownian limit in the NLD. Therefore, we can conclude that the condition imposed for FnGD in the NLD model also dictates that such systems necessarily revert to Brownian motion at large length scales. To be precise, the emergence of the Brownian behaviour can be observed for $x \gg x_0$. Whereas for the range of $x \lesssim x_0$, the system inherently exhibits strongly non-Gaussian dynamics althougth the mean-squared displacement is linear in time in any of these regimes.

\newpage
\section{Solutions to Exponential kernel}
The FnGD regime observed in molecular diffusion within various complex fluids using IQENS measurements follow a specific exponential kernel. In this section, we begin by formulating the kernel in 3D and show it matches with $I_{FnG}(Q,t)$ given in eq. \eqref{iqt_fngd}.

In 3D, we denote $\mathbf{x} \equiv \mathbf{r}$ and the corresponding length scale $r_0$. The jump-kernel is a function of only the radial coordinate $r \equiv |\mathbf{r}|$,
\begin{equation}
\label{exp kernel 3D}
f(r/r_0) = \frac{1}{4\pi} \frac{e^{-r/r_0}}{(r/r_0)}
\end{equation}
From this, we can calculate the radial Fourier transform, $\hat{f}(Qr_0)$ to be,
\begin{equation}
\label{FT exp kernel}
\hat{f}(Qr_0) = \frac{1}{1+(Qr_0)^2}
\end{equation}
Plugging this back into the solutions for NLD from eq. \eqref{NLD FT soln} with $d = 3$, we have the SISF given by,
\begin{equation}
I_s(Q,t) = I_0(Q) \exp \left[ - \frac{(Qr_0)^2}{1 + (Qr_0)^2} \frac{t}{\tau_j}\right]
\end{equation}
which precisely matches with the $I_{FnG}(Q,t)$ obtained in the FnGD limit of cage-jump diffusion model in eq. \eqref{iqt_fngd}. While we have presented the kernel in 3D, essentially, the Fourier-transform of the kernel in eq. \eqref{FT exp kernel} can be considered to be valid for any $d$-dimensional system with $d \le 4$. In fact, in eq. \eqref{kernel_gen}, the choice of $c_n = (-1)^n$ leads to a convergent sum ($|Qx_0|<1$) of the form given in eq. \eqref{FT exp kernel}.

\subsection{Complete solutions through subordination}
Here we present a solution to this kernel for any $d$-dimensional system, through the subordination technique that was established in the previous section. For the given exponential kernel, we have 
\begin{equation}
\label{exp jump kernel FT}
\hat{g}(Qx_0) = \frac{1}{1 + (Qx_0)^2} \frac{t}{\tau_j}
\end{equation}
Using the above in eq. \eqref{subordination connection}, to calculate the distribution of the subordinating process
\begin{equation}
T(\tau, t) = \mathcal{L}^{-1} \left\{ \exp \left[ -\frac{au}{1 + au} \frac{t}{\tau_j} \right] \right\}
\end{equation}
where we have substituted $Q^2 = u$ and $x_0^2 = a$. The above equation can be solved by expanding the exponential,
\begin{equation}
T(\tau, t) = e^{-t/\tau_j} \sum_{n=0}^\infty \frac{(t/\tau_j)^n}{n!} \mathcal{L}^{-1} \left\{ \left[ \frac{1}{1 + au} \right]^n \right\}
\end{equation} 
Solving the inverse Laplace transform of the terms in the summation and plugging back $a=x_0^2$, we have,
\begin{equation}
\label{exp kernel subordinator}
T(\tau, t) = e^{-t/\tau_j} \left[\delta(\tau) + \sum_{n=1}^\infty \left( \frac{t}{\tau_j} \right)^n \left( \frac{1}{x_0}\right)^{2n} \frac{\tau^{n-1} e^{-\tau/x_0^2} }{n! (n-1)!}  \right]
\end{equation}
The above equation provides the temporal variation of the distribution of the subordinating stochastic variable $\tau(t)$. The plot of this function indicates that for short times, it behaves like an exponential distribution, while at long times it eventually transforms into Gaussian centered around $\tau = D_j t$, with variance going to zero at long times. 

\vspace{0.5cm}
Before turning to short and long time behaviour of this function, we shall first look at the complete solution in terms of van-Hove correlation for a $d$-dimensional jump diffusion process governed by the kernel given in eq. \eqref{exp jump kernel FT}. In order to achieve this, we explicitly calculate the $G_s(\mathbf{x}, t)$ from eq. \eqref{subordination integral transform},
\begin{equation}
\begin{gathered}
G_s(\mathbf{x},t) = \int\displaylimits_{0}^\infty d\tau\; T(\tau, t)  \frac{ e^{-\mathbf{x}^2/(4\tau)} }  {(4\pi\tau)^{d/2}} \\
G_s(\mathbf{x},t) = e^{-t/\tau_j}   \left[\delta(\mathbf{x}) + \sum_{n=1}^\infty \left( \frac{t}{\tau_j} \right)^n \left( \frac{1}{x_0}\right)^{2n}  \frac{1}{n! (n-1)!} \int\displaylimits_{0}^\infty d\tau\; \tau^{n-1} e^{-\tau/x_0^2} \; \frac{ e^{-\mathbf{x}^2/(4\tau)} }  {(4\pi\tau)^{d/2}}  \right]  
\end{gathered}
\end{equation}
Solving for the integral, we obtain the complete solution for the jump-diffusion model, valid in $d$-dimensional systems ($d \le 4$),
\begin{equation}
\label{gsrt d-dimensional soln}
G_s(\mathbf{x},t) = e^{-t/\tau_j} \left[ \delta(\mathbf{x}) + \frac{2}{(2\pi x_0)^{d/2}} 
\sum_{n=1}^\infty \left( \frac{t}{\tau_j} \right)^n \left(\frac{|\mathbf{x}|}{2x_0}\right)^{n}  \frac{|\mathbf{x}|^{-d/2}}{n! (n-1)!} K_{n-\frac{d}{2}}\left( \frac{|\mathbf{x}|}{x_0} \right) \right]
\end{equation}
where $K_m(z)$ is the $m$-th order modified Bessel function of the second kind. Depending the dimension of the system, the behvaviour of the self-correlation function varies. In order to understand them in detail, we shall delve deeper into the short and long time limits.

\subsection{Limiting Behaviour of $G_s(\mathbf{x},t)$ at short and long times}
The short and long time limits of the dynamics in the system is decided by considering the typical jump timescale of the system, $\tau_j$. In order to understand the limiting behaviour we further simplify eq. \eqref{exp kernel subordinator} into a more tractable form by reducing the summation to modified Bessel function of first kind,
\begin{equation}
\label{exp kernel subordinator reduced}
T(\tau,t) = e^{-t/\tau_j} \left[ \delta(\tau) + e^{-\tau/x_0^2} \sqrt{\frac{1}{\tau x_0^2} \frac{t}{\tau_j}} I_1 \left(2 \sqrt{\frac{\tau}{x_0^2} \frac{t}{\tau_j}} \right)  \right]
\end{equation}
where $I_1(z)$ is the modified Bessel function of the first kind. Using this function, we can calculate the asymptotic limits of the subordinating kernel, $T(\tau, t)$. The short time asymptotic limit can be considered by $t \ll \tau_j$ or $t/\tau_j \ll 1$,
\begin{equation}
\label{exp kernel subordinator short time}
T^{st}(\tau, t) = e^{-t/\tau_j} \left[ \delta(\tau) + \left(\frac{t}{\tau_j}\right) \frac{1}{x_0^2}e^{-\tau/x_0^2} + \mathcal{O}\left( \left[\frac{t}{\tau_j}\right]^2 \right)  \right]
\end{equation}
Solving for $G_s(\mathbf{x},t)$ in the short-time asymptotic limit $t \ll \tau_j$, using the above subordinator, we have,
\begin{equation}
	\label{gsrt d-dimensional short time}
	G_s(\mathbf{x}, t) = e^{-t/\tau_j} \left( \delta(\mathbf{x}) + \frac{1}{(2\pi)^{d/2}} \frac{t}{\tau_j} \frac{x^{1-d/2}}{x_0^{1+d/2}} K_{1 - \frac{d}{2}} \left(\frac{|\mathbf{x}|}{x_0}\right) + \mathcal{O}\left( \left[\frac{t}{\tau_j}\right]^2 \right) \right)
\end{equation}
Ignoring the $\delta(\mathbf{x})$, which essentially correspond to the initial condition, we can observe the typical behaviour of leading order in $G_s(\mathbf{x}, t)$ for $d=1, 2, 3$,
\begin{equation}
\begin{gathered}
G_s^{1d}(x, t) = e^{-t/\tau_j} \left[\frac{t}{\tau_j} \frac{e^{-|x|/x_0}}{2x_0} + \mathcal{O}\left( \left[\frac{t}{\tau_j}\right]^2 \right)  \right] \\
G_s^{2d}(\mathbf{x}, t) = e^{-t/\tau_j} \left[\frac{t}{\tau_j} \frac{1}{2\pi} K_0 \left(\frac{|\mathbf{x}|}{x_0}\right)  + \mathcal{O}\left( \left[\frac{t}{\tau_j}\right]^2 \right)  \right] \\
G_s^{3d}(\mathbf{r}, t) = e^{-t/\tau_j} \left[\frac{t}{\tau_j}\frac{1}{4\pi r_0^2} \frac{e^{-r/r_0}}{r} + \mathcal{O}\left( \left[\frac{t}{\tau_j}\right]^2 \right)  \right] \\
\end{gathered}
\end{equation}
where we have categorically used $\mathbf{x} \equiv \mathbf{r}$ for $d=3$. The short-time behaviour of $G_s(\mathbf{x},t)$ is exponentially with respect to the displacement $\mathbf{x}$ in $d=1, 3$. Moreover, for $d=2$ to as well, it exhibits exponential feature when $|\mathbf{x}| \gg x_0$, as in this limit, $K_0(z) \rightarrow e^{-z}$. 

\vspace{0.5cm}
Therefore, we note that when the system is strongly non-Gaussian, the behaviour of the leading order in $G_s(\mathbf{x},t)$ typically follows an exponential form with $\mathbf{x}$.

\vspace{0.5cm}
In order to calculate the long-time behaviour, we consider $t \gg \tau_j$, by taking the limit $t/\tau_j \gg 1$ in eq. \eqref{exp kernel subordinator reduced},
\begin{equation}
\label{exp kernel subordinator long time}
T^{lt}(\tau, t) = \frac{1}{2 \tau^{3/4}} \left(\frac{1}{D_j t}\right)^{1/4} \frac{\exp \left[ \frac{\left(\sqrt{\tau/(D_jt)} - 1\right)^2}{\tau_j/t} \right]}{\sqrt{\pi (\tau_j/t)}}
\end{equation}
where we have used $D_j = x_0^2/\tau_j$. As mentioned earlier, this exhibits a Gaussian like behaviour for $\tau(t)$, with mean centered around $D_j t$. Upon further considering the limit $\tau_j/t \rightarrow 0$, the above equation reduces to,
\begin{equation}
\label{exp kernel subordinator long time 2}
T^{lt}(\tau, t) \simeq \frac{(D_j t)^{1/4}}{2 \tau^{3/4}} \delta\left( \sqrt{\tau} - \sqrt{D_j t} \right)
\end{equation}
Therefore, using the above limiting distribution in eq. \eqref{subordination integral transform}, we obtain,
\begin{equation}
\label{gsrt d-dimensional long time}
G_s^{lt}(\mathbf{x},t) = \frac{1}{(4\pi D_j t)^{d/2}} \exp \left[ - \frac{\mathbf{x}^2}{4D_j t}\right]
\end{equation}
which is exactly corresponds to the solution of $d$-dimensional Brownian motion. 

\vspace{0.5cm}

\textbf{Therefore, we conclude that the behaviour of $G_s(\mathbf{x},t)$ exhibits a typical exponential characteristic at timescales short compared to $\tau_j$, while at sufficiently long-times ($t \gg \tau_j$), Brownian motion is restored.}

\bibliography{MS_refs}